\documentstyle[prd,aps,preprint,tighten,epsfig]{revtex}

\begin{document}

\draft

\title{On the Koide-like Relations for the Running Masses \\
of Charged Leptons, Neutrinos and Quarks}
\author{{\bf Zhi-zhong Xing}\thanks{E-mail: xingzz@mail.ihep.ac.cn}
~ and ~ {\bf He Zhang}\thanks{E-mail: zhanghe@mail.ihep.ac.cn}}
\address{CCAST (World Laboratory), P.O. Box 8730,
Beijing 100080, China \\
and Institute of High Energy Physics,
Chinese Academy of Sciences, \\
P.O. Box 918, Beijing 100049, China}

\maketitle

\begin{abstract}
Current experimental data indicate that the Koide relation for the
{\it pole} masses of charged leptons, which can be parametrized as
$Q^{\rm pole}_l = 2/3$, is valid up to the accuracy of ${\cal
O}(10^{-5})$. We show that the {\it running} masses of charged
leptons fail in satisfying the Koide relation (i.e., $Q^{}_l(\mu)
\neq 2/3$), but the discrepancy between $Q^{}_l (\mu)$ and $Q^{\rm
pole}_l$ is only about $0.2\%$ at $\mu = M^{}_Z$. The Koide-like
relations for the running masses of neutrinos ($1/3 <
Q^{}_\nu(M^{}_Z) < 0.6$), up-type quarks ($Q^{}_{\rm U}(M^{}_Z)
\sim 0.89$) and down-type quarks ($Q^{}_{\rm D}(M^{}_Z) \sim
0.74$) are also examined from $M^{}_Z$ up to the typical seesaw
scale $M^{}_R \sim 10^{14}$ GeV, and they are found to be nearly
stable against radiative corrections. The approximate stability of
$Q^{}_{\rm U}(\mu)$ and $Q^{}_{\rm D}(\mu)$ is mainly attributed
to the strong mass hierarchy of quarks, while that of
$Q^{}_l(\mu)$ and $Q^{}_\nu(\mu)$ is essentially for the reason
that the lepton mass ratios are rather insensitive to radiative
corrections.
\end{abstract}

\newpage

\framebox{\Large\bf 1} ~
Thanks to the precise measurements of
three charged lepton masses \cite{PDG},
\begin{eqnarray}
m^{}_e & = & \left (0.510998918 \pm 0.000000044 \right ) ~ {\rm
MeV} \; , ~~
\nonumber \\
m^{}_\mu & = & \left (105.6583692 \pm 0.0000094 \right ) ~ {\rm
MeV} \; ,
\nonumber \\
m^{}_\tau & = & \left (1776.99^{+0.29}_{-0.26} \right ) ~ {\rm
MeV} \; ,
%       (1)
\end{eqnarray}
the empirical Koide mass relation \cite{Koide1}
\begin{eqnarray}
Q^{\rm pole}_l \; \equiv \; \frac{m^{}_e + m^{}_\mu + m^{}_\tau}{
\left (\sqrt{m^{}_e} + \sqrt{m^{}_\mu} + \sqrt{m^{}_\tau} \right
)^2} \; = \; \frac{2}{3}
%       (2)
\end{eqnarray}
has been testified up to the accuracy of ${\cal O}(10^{-5})$.
Namely, the above experimental data yield $-0.00001 \leq Q^{\rm
pole}_l - 2/3 \leq +0.00002$. This precision is so amazing that
there might exist some underlying physics behind the Koide
relation. A number of authors have tried to understand Eq. (2) and
extend it to neutrino masses or quark masses based on possible
flavor symmetries and phenomenological conjectures
\cite{Koide2,Foot,Li,Gerard}. It is certainly impossible to get a
universal Koide relation for both charged fermions and neutrinos
by using the experimental data at low energy scales, nor is it
likely to achieve such a mass relation at any high energy scales.
In this category of exercises, one has to take care of the
concepts of fermion masses and clarify the running mass behaviors
from one energy scale to another.

Note that Eq. (2) holds for the {\it pole} masses of three charged
leptons, which have been denoted as $m^{}_e$, $m^{}_\mu$ and
$m^{}_\tau$. Is it also applicable to the {\it running} masses
$m^{}_e(\mu)$, $m^{}_\mu(\mu)$ and $m^{}_\tau(\mu)$ at a given
energy scale (e.g., $\mu = M^{}_Z$)? We are going to answer this
question both analytically and numerically.

If the Koide relation is not a universal relation for fermion
masses at low energy scales, one may wonder whether a universal
Koide-like relation is achievable at a superhigh energy scale. We
shall show that the answer to this question is negative for both
charged fermions and neutrinos, but the dynamical reasons are
somehow different in these two cases.

Let us define the following Koide-like parameters for the running
masses of charged leptons ($f = l$), up-type quarks ($f = {\rm
U}$), down-type quarks ($f= {\rm D}$) and neutrinos ($f = \nu$):
\begin{eqnarray}
Q^{}_f (\mu) = \frac{m^{}_x (\mu) + m^{}_y (\mu) + m^{}_z
(\mu)}{\left [ \sqrt{m^{}_x (\mu)} + \sqrt{m^{}_y (\mu)} +
\sqrt{m^{}_z (\mu)} \right ]^2} \ ,
%       (3)
\end{eqnarray}
where
\begin{equation}
\left ( x, ~y, ~z \right ) \; =\; \left \{
\begin{array}{l}
\left ( e, ~\mu, ~\tau \right ) \; , ~~~~~~ {\rm for} ~ f=l \; ; \\
\left ( 1, ~2, ~3 \right ) \; , ~~~~~~ {\rm for} ~ f=\nu \; ; \\
\left ( u, ~c, ~t \right ) \; , ~~~~~~ {\rm for} ~ f={\rm U} \; ; \\
\left ( d, ~s, ~b \right ) \; , ~~~~~~ {\rm for} ~ f={\rm D} \; .
\end{array}
\right . ~~
%       (4)
\end{equation}
As pointed out by G$\rm\acute{e}$rard {\it et al} \cite{Gerard},
$Q^{}_f = 1$ (maximum) would hold when three fermion masses had
the extremely hierarchical spectrum (e.g., $m^{}_x \ll m^{}_y \ll
m^{}_z$), while $Q^{}_f = 1/3$ (minimum) would occur if three
fermion masses were exactly degenerate (i.e., $m^{}_x = m^{}_y =
m^{}_z$). It might be a puzzle that $Q^{\rm pole}_l$ happens to
take the mean value of $Q^{\rm min}_f$ and $Q^{\rm max}_f$. The
main purpose of this paper is just to examine the deviations of
realistic $Q^{}_l$, $Q^{}_\nu$, $Q^{}_{\rm U}$ and $Q^{}_{\rm D}$
from $Q^{\rm pole}_l$ at and above the electroweak scale.

\vspace{0.4cm}

\framebox{\Large\bf 2} ~
First of all, let us calculate the
discrepancy between $Q^{}_l(M^{}_Z)$ and $Q^{\rm pole}_l$. The
running masses of three charged leptons $m^{}_l (\mu)$ are related
with their corresponding pole masses $m^{}_l$ in a very simple way
\cite{running},
\begin{eqnarray}
m^{}_l(\mu) \; = \; m^{}_l \left (1 - \Delta^{}_l \right ) \ ,
%       (5)
\end{eqnarray}
where
\begin{eqnarray}
\Delta^{}_l = \frac{\alpha(\mu)}{\pi} \left [ \frac{3}{2} \ln
\frac{\mu}{m^{}_l(\mu)} + 1 \right ]
%       (6)
\end{eqnarray}
with $\alpha(\mu)$ being the fine-structure constant at the energy
scale $\mu$. It is then straightforward to establish the
relationship between $Q^{}_l (\mu)$ and $Q^{\rm pole}_l$ by
combining Eqs. (3) and (5). Taking account of the strong mass
hierarchy $m^{}_e \ll m^{}_{\mu} \ll m^{}_{\tau}$ (more
explicitly, $m^{}_e/m^{}_\mu \approx 0.00484$ and
$m^{}_\mu/m^{}_\tau \approx 0.0595$ \cite{PDG}), we find that it
is more instructive to make the following analytical
approximation:
\begin{eqnarray}
Q^{}_l (\mu) \simeq Q^{\rm pole}_l \left [ 1 +
\sqrt{\frac{m^{}_{\mu}}{m^{}_{\tau}}} \left ( \Delta^{}_{\mu} -
\Delta^{}_{\tau} \right ) \right] \ .
%       (7)
\end{eqnarray}
One can clearly see that the deviation of $Q^{}_l (\mu)$ from
$Q^{\rm pole}_l$ is strongly suppressed, due to the smallness of
$\sqrt{m^{}_{\mu}/m^{}_{\tau}}$ and that of $\Delta^{}_l$.

At $\mu = M^{}_Z$ with $\alpha(M^{}_Z)=(128.89)^{-1}$
\cite{alpha}, the running masses of three charged leptons can be
evaluated by solving Eq. (5) with the inputs of the pole masses
given in Eq. (1). For our purpose, we only need to make use of the
central values of $m^{}_e$, $m^{}_\mu$ and $m^{}_\tau$ in the
calculation. The results are
\begin{eqnarray}
m^{}_e(M^{}_Z) & = & 0.486755106 ~{\rm MeV} \ , \nonumber \\
m^{}_\mu(M^{}_Z) & = & 102.740394 ~{\rm MeV} \ , \nonumber \\
m^{}_\tau(M^{}_Z) & = & 1746.56 ~{\rm MeV} \ .
%       (8)
\end{eqnarray}
With the help of Eq. (6), we obtain $\Delta^{}_\mu = 0.0276$ and
$\Delta^{}_\tau = 0.0171$ at $M^{}_Z$. Therefore,
\begin{equation}
Q^{}_l(M^{}_Z) = 1.00188 \times Q^{\rm pole}_l = 0.66792 \; ,
%       (9)
\end{equation}
achieved from Eqs. (3) and (5). If the analytical approximation
made in Eq. (7) is used to estimate $Q^{}_l (M^{}_Z)$, one can get
$Q^{}_l(M^{}_Z) \approx 1.00256 \times Q^{\rm pole}_l = 0.66837$.
This result is compatible with Eq. (9) and implies that Eq. (7) is
actually a reasonable approximation.

Obviously, the discrepancy between $Q^{}_l(M^{}_Z)$ and $Q^{\rm
pole}_l$ is only about $0.2\%$. This tiny difference makes sense
in physics, because it is much larger than the accuracy of Eq.
(2). In other words, the possibility of $Q^{}_l(M^{}_Z) = 2/3$,
which inversely leads to $Q^{\rm pole}_l - 2/3 = -0.00188$, has
been ruled out by the experimental data. We are therefore left
with the conclusion that the running and pole masses of three
charged leptons cannot simultaneously satisfy the Koide relation.

\vspace{0.4cm}

\framebox{\Large\bf 3} ~
Now let us turn to the Koide-like
relations of quark masses. We concentrate on the running masses of
six quarks at $\mu = M^{}_Z$ \cite{qmass}
%%%%%%%%%%%%%%%%%%%%%%%
\footnote{We do not consider the pole masses of six quarks,
because they are neither directly measurable nor relevant for
model building. In particular, the pole masses of three light
quarks ($u,d,s$) involve large uncertainties, since they can only
be evaluated in the region with a large $\alpha^{}_s (\mu)$
\cite{update1}.},
%%%%%%%%%%%%%%%%%%%%%%%
\begin{eqnarray}
m^{}_u(M^{}_Z) & = & \left (1.7 \pm 0.4 \right )
~{\rm MeV} \ ,
\nonumber \\
m^{}_c(M^{}_Z) & = & \left (0.62 \pm 0.03 \right ) ~{\rm GeV} \ ,
~~~~~~
\nonumber \\
m^{}_t(M^{}_Z) & = & \left (171 \pm 3 \right ) ~{\rm GeV} \ ;
%       (10)
\end{eqnarray}
and
\begin{eqnarray}
m^{}_d(M^{}_Z) & = & \left (3.0 \pm 0.6 \right ) ~{\rm MeV} \ ,
\nonumber \\
m^{}_s(M^{}_Z) & = & \left (54 \pm 11 \right ) ~{\rm MeV} \ ,
\nonumber \\
m^{}_b(M^{}_Z) & = & \left (2.87 \pm 0.03 \right ) ~{\rm GeV} \ .
%       (11)
\end{eqnarray}
The Koide-like parameters $Q^{}_{\rm U}$ and $Q^{}_{\rm D}$
defined in Eq. (3) can then be calculated at $M^{}_Z$ with the
help of Eqs. (10) and (11). For simplicity, only the central
values of those quark masses are taken into account. The results
are
\begin{eqnarray}
Q^{}_{\rm U}(M^{}_Z) \; \approx \; 0.89 \ , ~~~~~~ Q^{}_{\rm
D}(M^{}_Z) \; \approx \; 0.74 \ .
%       (12)
\end{eqnarray}
We see that both numbers significantly deviate from $2/3$ at the
electroweak scale.

To examine whether the Koide-like relation $Q^{}_{\rm U}(M^{}_X)
\approx Q^{}_{\rm D} (M^{}_X) \approx 2/3$ could hold at a
superhigh energy scale $M^{}_X$ (e.g., $M^{}_X \sim 10^{14 - 16}$
GeV), one may make use of the one-loop renormalization-group
equations (RGEs) of quark Yukawa couplings \cite{RGE} in the
standard model (SM) or in the minimal supersymmetric standard
model (MSSM). Note that the strong hierarchy of charged fermion
masses and that of the quark mixing angles allow us to simplify
those RGEs to a great extent \cite{ratios}. In particular, the RGE
evolution of $m^{}_u/m{}_c$, $m^{}_d/m{}_s$ and $m^{}_e/m{}_\mu$
from $M^{}_Z$ to $M^{}_X$ are negligibly small in both the SM and
the MSSM. Radiative corrections to the mass ratios
$m^{}_c/m^{}_t$, $m^{}_s/m^{}_b$ and $m^{}_\mu/m^{}_\tau$ can be
written as
\begin{eqnarray}
\frac{m^{}_c (M^{}_X)}{m^{}_t (M^{}_X)} & \approx & \frac{m^{}_c
(M^{}_Z)}{m^{}_t (M^{}_Z)} \chi^{}_{\rm U} \ ,
\nonumber \\
\frac{m^{}_s (M^{}_X)}{m^{}_b (M^{}_X)} & \approx & \frac{m^{}_s
(M^{}_Z)}{m^{}_b (M^{}_Z)} \chi^{}_{\rm D} \ ,
\nonumber \\
\frac{m^{}_\mu (M^{}_X)}{m^{}_\tau (M^{}_X)} & \approx &
\frac{m^{}_\mu (M^{}_Z)}{m^{}_\tau (M^{}_Z)} \chi^{}_l \ ,
%       (13)
\end{eqnarray}
where the evolution functions $\chi^{}_{\rm U}$, $\chi^{}_{\rm D}$
and $\chi^{}_l$ are defined by
\begin{eqnarray}
\chi^{}_f \equiv \exp\left[ \frac{1}{16\pi^2} \int^{\ln
M^{}_X}_{\ln M^{}_Z} \left( a^{}_f y^2_t + b^{}_f y^2_b + c^{}_f
y^2_\tau \right){\rm d} t \right]
%       (14)
\end{eqnarray}
with $t =\ln(\mu)$, $f=({\rm U}, {\rm D},l)$ and $y^{}_i$
($i=t,b,\tau$) being the Yukawa couplings. In the SM, $a^{}_{\rm
U} = - b^{}_{\rm U} = - a^{}_{\rm D} = b^{}_{\rm D} = c^{}_l =
-3/2$ and $c^{}_{\rm U} = c^{}_{\rm D} = a^{}_l = b^{}_l =0$ hold;
while in the MSSM, we have $a^{}_{\rm U} = b^{}_{\rm D} = c^{}_l =
-3$, $b^{}_{\rm U} = a^{}_{\rm D} =-1$ and $c^{}_{\rm U} =
c^{}_{\rm D} = a^{}_l = b^{}_l =0$. The running behaviors of
$\chi^{}_f$ are typically illustrated in Fig. 1. One can see that
$\chi^{}_l$ is insensitive to radiative corrections in both the SM
and the MSSM. In contrast, $\chi^{}_{\rm U}$ may significantly
depart from 1 when $M^{}_X \gg M^{}_Z$ holds, especially for
sizable $\tan\beta$ in the MSSM. Note that $\chi^{}_{\rm D} > 1$
holds in the SM, as a consequence of $a^{}_{\rm D} >0$ in Eq.
(14).

With the help of $\chi^{}_f$ (for $f= {\rm U}, {\rm D}, l$) given
above, it is easy to derive the Koide-like parameter $Q^{}_f
(M^{}_X)$ in terms of $Q^{}_f (M^{}_Z)$. We approximately obtain
\begin{eqnarray}
Q^{}_{\rm U} (M^{}_{X}) & \approx & Q^{}_{\rm U} (M^{}_Z) \left[ 1
+ 2\left(1- \sqrt{\chi^{}_{\rm U}} \right) \sqrt{\frac{m^{}_c
(M^{}_Z)}{m^{}_t (M^{}_Z)}} \right] \ ,
\nonumber \\
Q^{}_{\rm D} (M^{}_{X}) & \approx & Q^{}_{\rm D} (M^{}_Z) \left[ 1
+ 2\left(1- \sqrt{\chi^{}_{\rm D}} \right) \sqrt{\frac{m^{}_s
(M^{}_Z)}{m^{}_b (M^{}_Z)}} \right] \ ,
\nonumber \\
Q^{}_l (M^{}_{X}) & \approx & Q^{}_l (M^{}_Z) \left[ 1 + 2\left(1-
\sqrt{\chi^{}_l} \right) \sqrt{\frac{m^{}_\mu (M^{}_Z)}{m^{}_\tau
(M^{}_Z)}} \right] \ .
%       (15)
\end{eqnarray}
One can see that the deviation of $Q^{}_f (M^{}_X)$ from $Q^{}_f
(M^{}_Z)$ is significantly suppressed, because of (i) the strong
mass hierarchies of charged fermions and (ii) the small departure
of $\chi^{}_f$ from 1. Indeed, $\sqrt{m^{}_c/m^{}_t} \approx
0.06$, $\sqrt{m^{}_s/m^{}_b} \approx 0.14$ and $\sqrt{m^{}_\mu/
m^{}_\tau} \approx 0.24$ at $M^{}_Z$. Taking $M^{}_X = M^{}_R \sim
10^{14}$ GeV (the typical scale of heavy right-handed Majorana
neutrinos in the seesaw models \cite{SS}), for example, we get
$\sqrt{\chi^{}_{\rm U}} \approx 0.97$, $\sqrt{\chi^{}_{\rm D}}
\approx 1.03$ and $\sqrt{\chi^{}_{l}} \approx 1.00$ in the SM; or
$\sqrt{\chi^{}_{\rm U}} \approx 0.86$, $\sqrt{\chi^{}_{\rm D}}
\approx 0.91$ and $\sqrt{\chi^{}_{l}} \approx 0.98$ in the MSSM
with $\tan\beta = 50$. Therefore,
\begin{equation}
Q^{}_f (M^{}_R) - Q^{}_f (M^{}_Z) \; \approx \; \left \{
\begin{array}{l}
3.3\times10^{-3} \; , ~~~~~ {\rm for} ~ f = {\rm U} \; , \\
-6.5\times10^{-3} \; , \;\;\; {\rm for} ~ f = {\rm D} \; , \\
0 \; , ~~~~~~~~~~~~~~~~~ {\rm for} ~ f = l \;
\end{array}
\right .
%       (16)
\end{equation}
in the SM; or
%%%%%%%%%%%%%%%%%%%%%
\footnote{The deviation of $Q^{}_f (M^{}_R)$ from $Q^{}_f
(M^{}_Z)$ in the MSSM will become much milder, if $\tan\beta$
takes smaller values ($\tan\beta = 10$ shown in Fig. 1(b), for
example).}
%%%%%%%%%%%%%%%%%%%%%
\begin{equation}
Q^{}_f (M^{}_R) - Q^{}_f (M^{}_Z) \; \approx \; \left \{
\begin{array}{l}
1.4\times10^{-2} \; , ~~~~ {\rm for} ~ f = {\rm U} \; , \\
1.8\times10^{-2} \; , ~~~~ {\rm for} ~ f = {\rm D} \; , \\
4.8\times10^{-3} \; , ~~~~ {\rm for} ~ f = l \;
\end{array}
\right .
%       (17)
\end{equation}
in the MSSM with $\tan\beta = 50$. It is obvious that radiative
corrections to the Koide-like parameters are very small.

\vspace{0.4cm}

\framebox{\Large\bf 4} ~ Finally, we consider the Koide-like
relation in the neutrino sector. Although the pole masses of three
neutrinos are different from their running masses at $\mu =
M^{}_Z$, this difference is negligibly tiny because it is strongly
suppressed by the Fermi coupling constant. For simplicity, we
mainly calculate the Koide-like parameter $Q^{}_\nu (M^{}_Z)$ and
examine its sensitivity to radiative corrections from $M^{}_Z$ up
to the seesaw scale $M^{}_R$.

Note that the absolute values of three neutrino masses remain
unknown, but their upper bound is expected to be $m^{}_i < 0.23
~{\rm eV}$ (for $i=1,2,3$) \cite{Mohapatra}. A global analysis of
current experimental data on neutrino oscillations yields
\cite{Vissani}: $\Delta m^{2}_{21} \equiv m^2_2 - m^2_1 =(8.0 \pm
0.3) \times 10^{-5}~ {\rm eV^2}$ and $\Delta m^{2}_{32} \equiv
m^2_3 - m^2_2 = \pm (2.5 \pm 0.3) \times 10^{-3} ~{\rm eV^2}$ at
the $99\%$ confidence level. Since the sign of $\Delta m^2_{32}$
has not been fixed, the neutrino mass spectrum can be classified
into two general categories:
\begin{itemize}
\item Normal hierarchy: $m^{}_1 < m^{}_2 < m^{}_3$, including the
possibility that three neutrino masses are nearly degenerate
($\Delta m^2_{32} > 0$).
\item Inverted hierarchy: $m^{}_3 <
m^{}_1 < m^{}_2$, including the possibility that three neutrino
masses are nearly degenerate ($\Delta m^2_{32} < 0$).
\end{itemize}
Allowing $m^{}_1$ or $m^{}_3$ to vary up to its upper bound and
inputting the experimental values of $\Delta m^2_{21}$ and $\Delta
m^2_{32}$, we may use Eq. (3) to calculate $Q^{}_\nu (M^{}_Z)$.
Our numerical results are illustrated in Fig. 2. One can see that
the upper limit of $Q^{}_\nu (M^{}_Z)$ is achieved at $m^{}_1 =0$
for the normal neutrino mass hierarchy or at $m^{}_3 =0$ for the
inverted neutrino mass hierarchy. Indeed,
\begin{equation}
\frac{1}{Q^{}_\nu (M^{}_Z)} \; =\; \left \{
\begin{array}{l}
1 + 2 \displaystyle \frac{\sqrt{\sqrt{\Delta m^2_{21}}
\sqrt{\Delta m^2_{21} + | \Delta m^2_{32} |}}}{\sqrt{\Delta
m^2_{21}} + \sqrt{\Delta m^2_{21} + | \Delta m^2_{32} |}} \; ,
~~~~~ (m^{}_1 \rightarrow 0) \; , \\ \\
1 + 2 \displaystyle \frac{\sqrt{\sqrt{|\Delta m^2_{32}|}
\sqrt{|\Delta m^2_{32}| - \Delta m^2_{21}}}}{\sqrt{|\Delta
m^2_{32}|} + \sqrt{|\Delta m^2_{32} - \Delta m^2_{21}}} \; , ~~~~
(m^{}_3 \rightarrow 0) \; .
\end{array}
\right .
%       (18)
\end{equation}
It is then straightforward to understand $Q^{}_\nu (M^{}_Z) \sim
0.6$ for $m^{}_1 \rightarrow 0$ and $Q^{}_\nu (M^{}_Z) \approx
0.5$ for $m^{}_3 \rightarrow 0$ (as shown in Fig. 2) in the
approximation of $|\Delta m^2_{32}| \gg \Delta m^2_{21}$. When
three neutrino masses are nearly degenerate, $Q^{}_\nu (M^{}_Z)$
approaches its minimal value $1/3$ no matter whether the sign of
$\Delta m^2_{32}$ is positive or negative.

The running masses of three neutrinos at $M^{}_R$ can be evaluated
by using the one-loop RGEs given in Ref. \cite{RGE}. In order to
understand the sensitivity of $Q^{}_\nu (M^{}_R)$ to radiative
corrections, it is more convenient to consider the RGE running
behaviors of $m^{}_i/m^{}_j$ (for $i\neq j$). We obtain
\begin{equation}
\frac{m^{}_i (M^{}_X)}{m^{}_j (M^{}_X)} \; \approx \; \frac{m^{}_i
(M^{}_Z)}{m^{}_j (M^{}_Z)} \chi^{(ij)}_\nu
%       (19)
\end{equation}
in the approximation of $\tau$-lepton dominance \cite{Xing06},
where
%%%%%%%%%%%%%%%%
\footnote{This formula is valid for Majorana neutrinos. If
neutrinos were Dirac particles, the coefficient $C/(8\pi^2)$ on
the right-hand side of Eq. (20) should be replaced by
$C/(16\pi^2)$ \cite{Xing06}.}
%%%%%%%%%%%%%%%%
\begin{eqnarray}
\chi^{(ij)}_\nu \equiv \exp\left[ \frac{C}{8\pi^2} \int^{\ln
M^{}_X}_{\ln M^{}_Z} y^2_\tau \left( |U^{}_{3i}|^2 - |U^{}_{3j}|^2
\right ){\rm d} t \right]
%       (20)
\end{eqnarray}
with $C = -1.5$ in the SM or $C = 1$ in the MSSM and $U_{3i}$ (for
$i=1,2,3$) being the elements of the $3\times 3$ lepton flavor
mixing matrix. Because of $y^2_\tau/(8\pi^2) \approx 1.3 \times
10^{-6}$ (SM) or $y^2_\tau/(8\pi^2) \approx 1.3 \times 10^{-6}
\left ( 1 + \tan^2\beta \right )$ (MSSM) at $M^{}_Z$, together
with $|U^{}_{3i}|^2 < 1$, the departure of $\chi^{(ij)}_\nu$ from
1 is expected to be tiny. This observation is illustrated in Fig.
3, where $\theta^{}_\nu \approx 33.8^\circ$ and $\theta \approx
45^\circ$ \cite{Xing06} have typically been input for the matrix
elements $|U^{}_{31}| = \sin\theta^{}_\nu \sin\theta$,
$|U^{}_{32}| = \cos\theta^{}_\nu \sin\theta$ and $|U^{}_{33}| =
\cos\theta$.

Note that $\chi^{(ij)}_\nu \approx 1$ is essentially independent
of the possible mass hierarchies of three neutrinos \cite{Zhang}.
It turns out that $Q^{}_\nu (M^{}_X) \approx Q^{}_\nu (M^{}_Z)$ is
a very good approximation. In view of Fig. 2, we conclude that
there is no hope to achieve the Koide-like relation $Q^{}_\nu
(\mu) \approx 2/3$ for neutrino masses.

\vspace{0.4cm}

\framebox{\Large\bf 5} ~ In summary, the updated Koide relation
for the pole masses of charged leptons satisfies $Q^{\rm pole}_l =
2/3$ at the precision level of ${\cal O}(10^{-5})$. This amazing
accuracy motivates us to examine whether the running masses of
charged fermions and neutrinos have the similar Koide relation at
a given energy scale. We have shown that the {\it running} masses
of charged leptons cannot satisfy the Koide relation (i.e.,
$Q^{}_l (\mu) \neq 2/3$), but its discrepancy from $Q^{\rm
pole}_l$ is only about $0.2\%$ at $\mu = M^{}_Z$. The Koide-like
relations for the running masses of neutrinos ($1/3 <
Q^{}_\nu(M^{}_Z) < 0.6$), up-type quarks ($Q^{}_{\rm U}(M^{}_Z)
\sim 0.89$) and down-type quarks ($Q^{}_{\rm D}(M^{}_Z) \sim
0.74$) are analyzed from $M^{}_Z$ up to the typical seesaw scale
$M^{}_R \sim 10^{14}$ GeV. We find that they are nearly stable
against radiative corrections, just like $Q^{}_l (\mu)$. The
approximate stability of $Q^{}_{\rm U}(\mu)$ and $Q^{}_{\rm
D}(\mu)$ are mainly attributed to the strong mass hierarchy of
quarks. In contrast, the approximate stability of $Q^{}_l (\mu)$
and $Q^{}_{\nu}(\mu)$ is essentially for the reason that the
lepton mass ratios are very insensitive to radiative corrections.

Although it is impossible to get a universal Koide relation for
both charged fermions and neutrinos at a given energy scale, our
work remains useful for model building in order to explore the
underlying similarities and differences between lepton and quark
masses. For example,
\begin{equation}
Q^{}_{\rm U} (M^{}_Z) > Q^{}_{\rm D} (M^{}_Z) > Q^{}_l (M^{}_Z)
> Q^{}_\nu (M^{}_Z)
%       (21)
\end{equation}
is an interesting result of our numerical analysis. Once the
absolute scale of neutrino masses is measured and the value of
$Q^{}_\nu (M^{}_Z)$ is fixed, one may then speculate whether these
four parameters could be related with one another in a
phenomenological way. On the other hand, we remark that more
theoretical efforts are needed to look for possible flavor
symmetries and their breaking effects behind the Koide-like
relations.

\acknowledgments{We would like to thank X.D. Ji, S. Zhou, S.H. Zhu
and S.L. Zhu for useful discussions. This work is supported in
part by the National Natural Science Foundation of China.}

\newpage

%%%%%%%%%%%%%%%%%%%% Fig. 1a %%%%%%%%%%%%%%%%
\begin{figure}
\vspace{3.5cm}
\epsfig{file=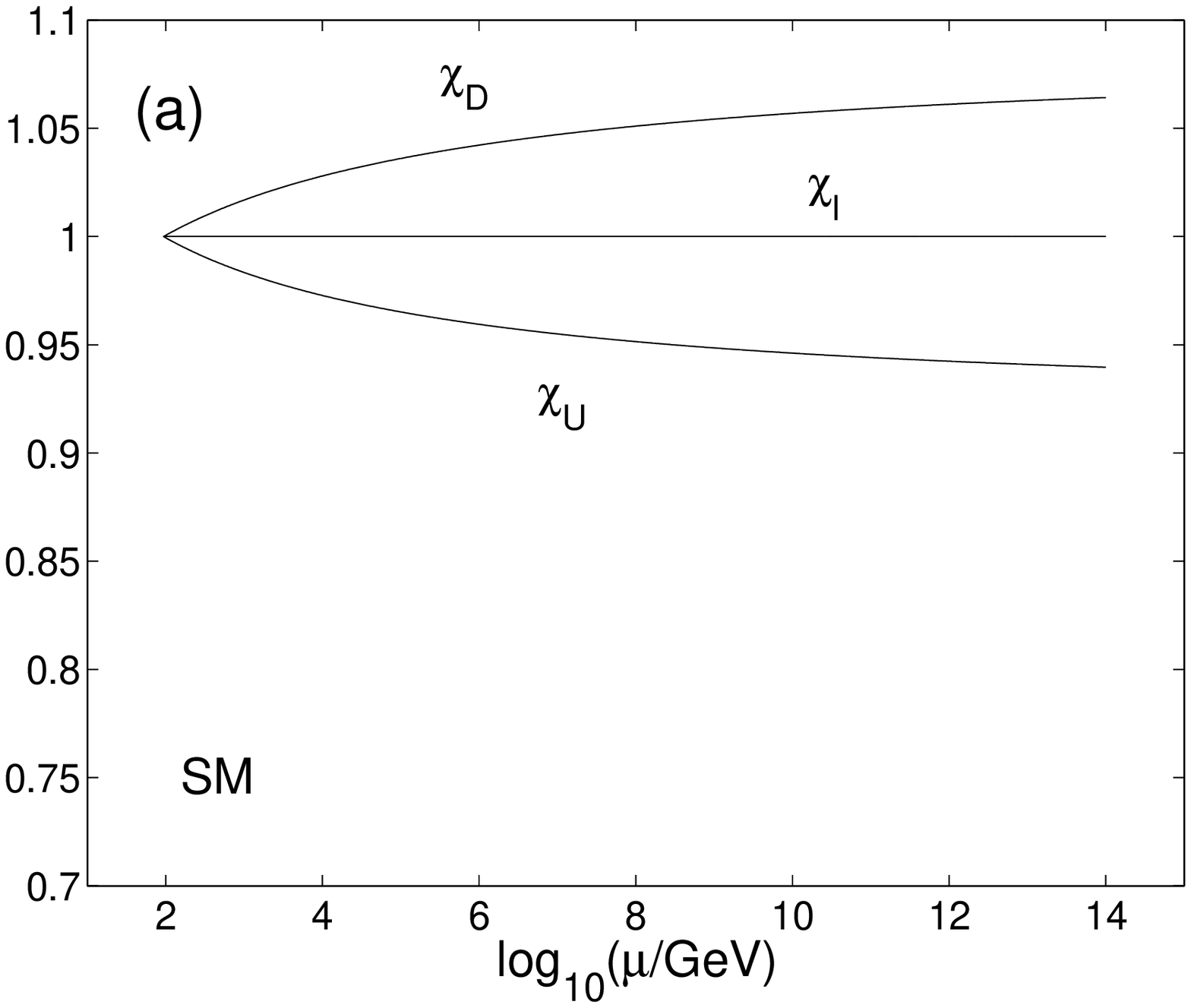,bbllx=-7cm,bblly=9cm,bburx=-3cm,bbury=13cm,%
width=1.8cm,height=1.8cm,angle=0,clip=0}\vspace{4.1cm}
\end{figure}
%%%%%%%%%%%%%%%%%%%% Fig. 1b %%%%%%%%%%%%%%%%
\begin{figure}
\epsfig{file=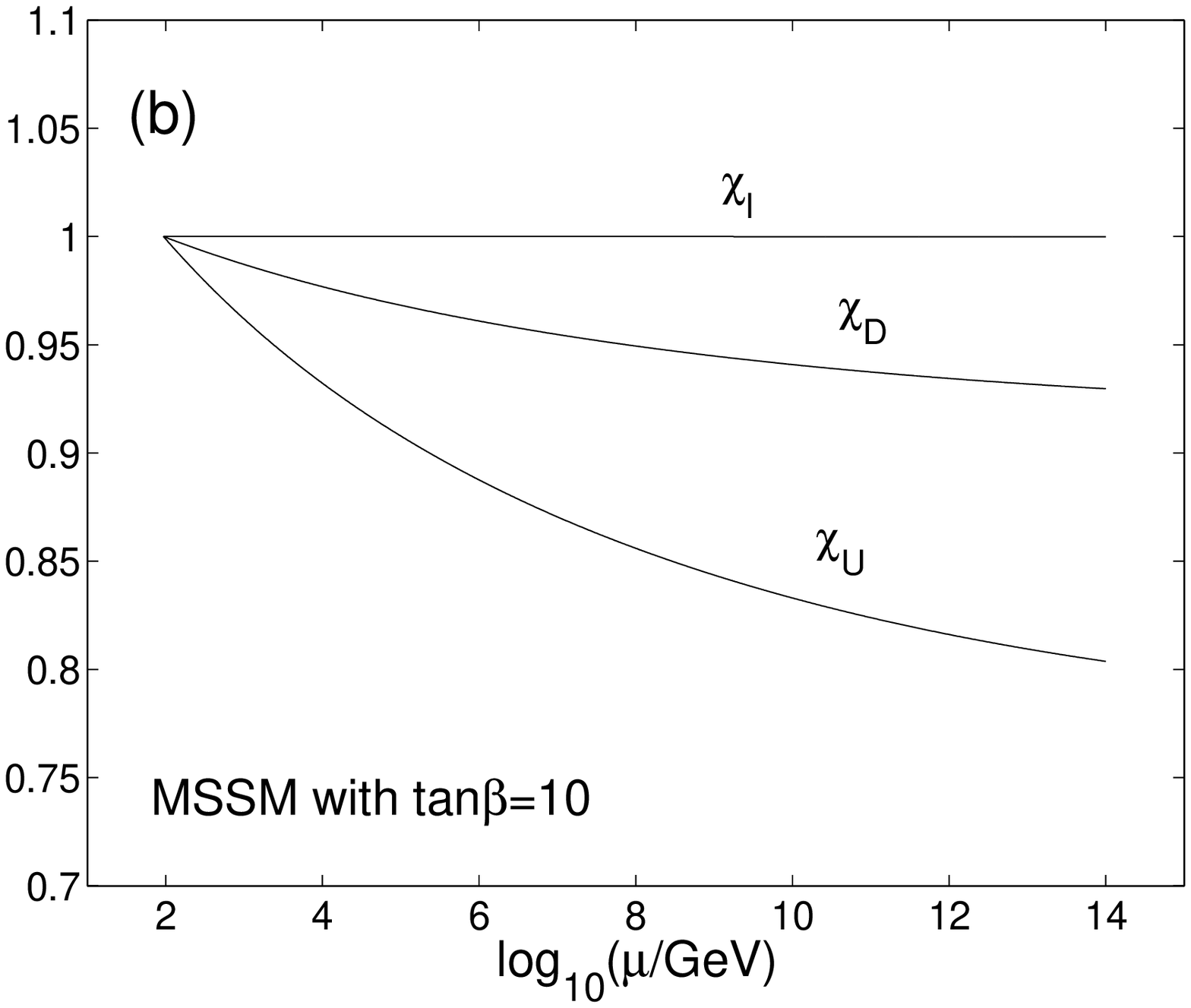,bbllx=-7cm,bblly=8.8cm,bburx=-3cm,bbury=12.8cm,%
width=1.8cm,height=1.8cm,angle=0,clip=0}\vspace{4cm}
\end{figure}
%%%%%%%%%%%%%%%%%%%% Fig. 1c %%%%%%%%%%%%%%%%
\begin{figure}
\epsfig{file=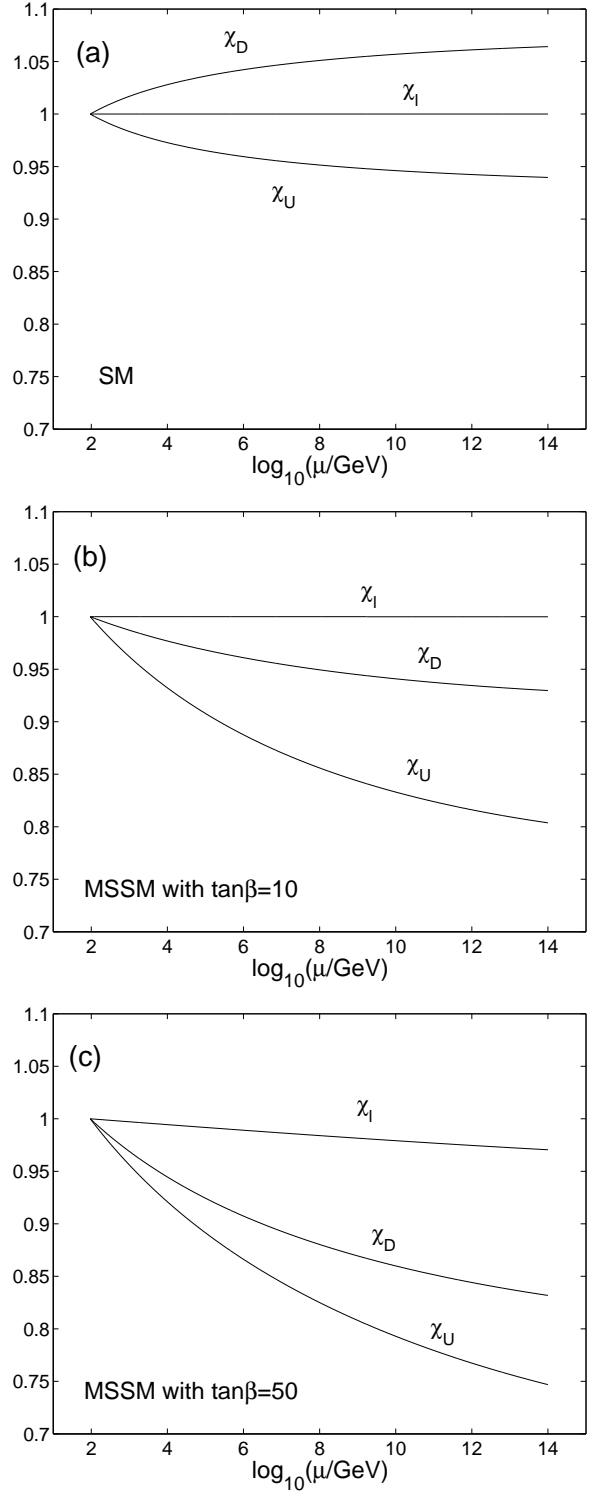,bbllx=-7cm,bblly=8.8cm,bburx=-3cm,bbury=12.8cm,%
width=1.8cm,height=1.8cm,angle=0,clip=0}
\vspace{1.5cm}
\caption{Illustration of $\chi^{}_{\rm U}$, $\chi^{}_{\rm D}$ and
$\chi^{}_l$ changing with the energy scale from $M^{}_Z$ to
$M^{}_R$.}\vspace{6cm}
\end{figure}
%%%%%%%%%%%%%%%%%%%%%%%%%%%%%%%%%%%%%%%%%%%

\newpage

%%%%%%%%%%%%%%%%%%%% Fig. 2 %%%%%%%%%%%%%%%%
\begin{figure}
%\begin{center}
\vspace{2cm}
\epsfig{file=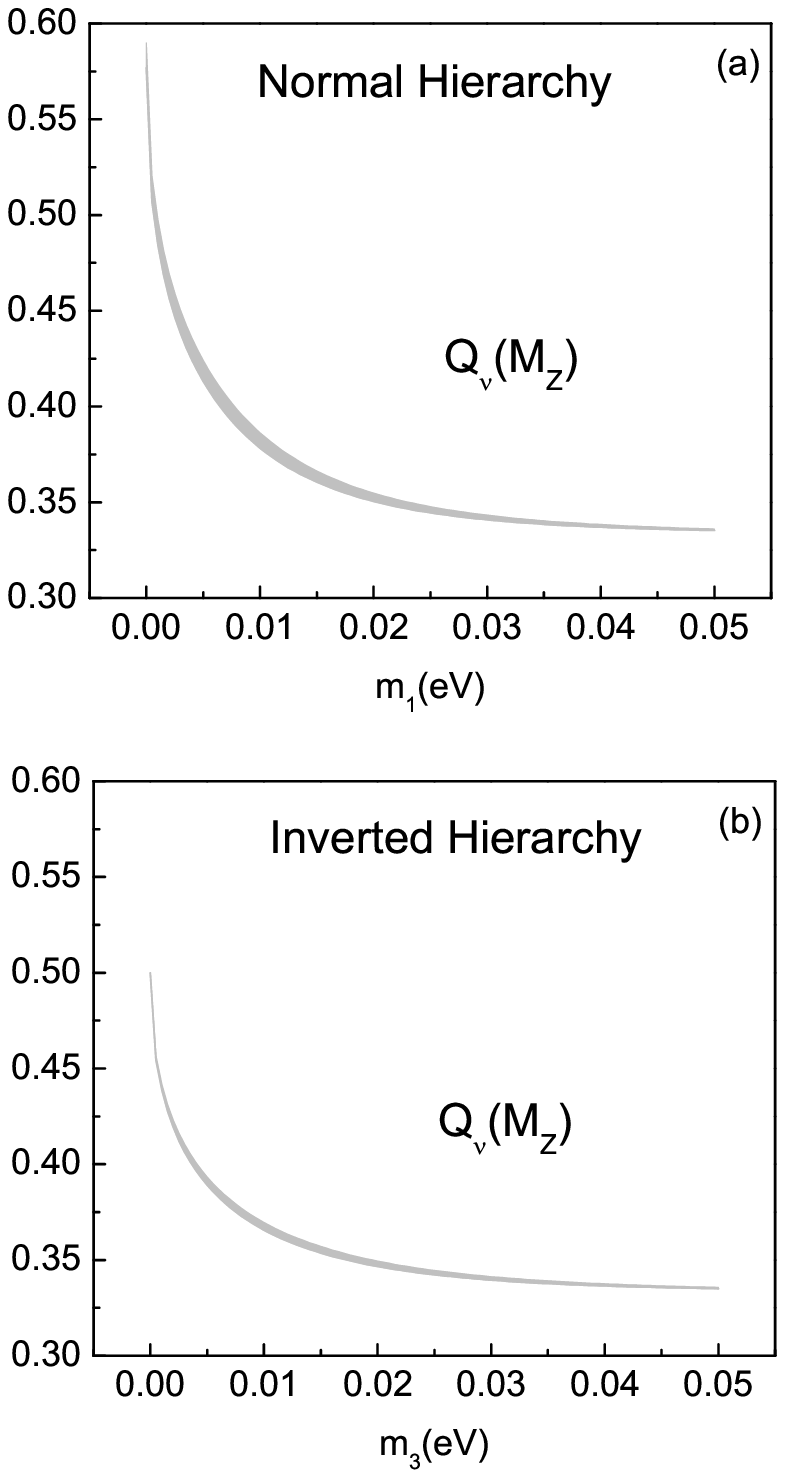,bbllx=3cm,bblly=21cm,bburx=7cm,bbury=25cm,%
width=4cm,height=4cm,angle=0,clip=0} \vspace{11.5cm} \caption{The
Koide-like parameter $Q^{}_\nu (M^{}_Z)$ evaluated by using
current neutrino oscillation data.}
%\end{center}
\end{figure}
%%%%%%%%%%%%%%%%%%%%%%%%%%%%%%%%%%%%%%%%%%%%

\newpage

%%%%%%%%%%%%%%%%%%%% Fig. 3 %%%%%%%%%%%%%%%%
\begin{figure}
\vspace{3.5cm}
\epsfig{file=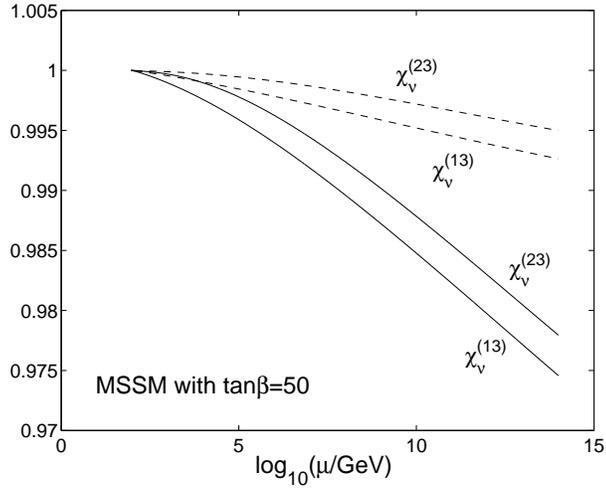,bbllx=-7cm,bblly=18cm,bburx=-3cm,bbury=22cm,%
width=1.8cm,height=1.8cm,angle=0,clip=0} \vspace{5.8cm}
\caption{Illustration of $\chi^{(ij)}_\nu$ changing with the
energy scale from $M^{}_Z$ to $M^{}_R$. The solid and dashed lines
denote the Majorana and Dirac cases, respectively. We have
typically input $\tan\beta = 50$ and $m^{}_1 = 0.2$ eV with
$m^2_{32}
>0$ in our calculation. The almost identical result can be
obtained when $m^2_{32} < 0$ is taken. Note that the deviation of
$\chi^{(ij)}_\nu$ from 1 will be much milder, if $\tan\beta$ takes
smaller values.}
\end{figure}
%%%%%%%%%%%%%%%%%%%%%%%%%%%%%%%%%%%%%%%%%%%%

\end{document}